\documentclass[a4paper,prl,twocolumn,superscriptaddress,showkeys,floatfix]{revtex4}
\usepackage[dvips]{graphicx}
\usepackage{amsmath}
\usepackage{amsthm}
\usepackage{amssymb}

\newcommand{\mean}[1]{\left\langle #1 \right\rangle}

\newcommand{\dd}{\mathrm{d}}
\newcommand{\erg}{\mathcal U}
\newcommand{\heat}{\mathcal C}
\begin{document}

\title{Finite size scaling of the 5D Ising model with free boundary conditions}

\date{\today} 

\author{P. H. Lundow} 
\email{per.hakan.lundow@math.umu.se} 


\author{K. Markstr\"om}
\email{klas.markstrom@math.umu.se} 

\affiliation{ Department of mathematics and mathematical statistics,
  Ume\aa{} University, SE-901 87 Ume\aa, Sweden}

\begin{abstract}
  There has been a long running debate on the finite size scaling for
  the Ising model with free boundary conditions above the upper
  critical dimension, where the standard picture gives a $L^2$ scaling
  for the susceptibility and an alternative theory has promoted a
  $L^{5/2}$ scaling, as would be the case for cyclic boundary.  In
  this paper we present results from simulation of the far largest
  systems used so far, up to side $L=160$ and find that this data
  clearly supports the standard scaling.  Further we present a
  discussion of why rigorous results for the random-cluster model
  provides both supports the standard scaling picture and provides a
  clear explanation of why the scalings for free and cyclic boundary
  should be different.
\end{abstract}


\maketitle

\section{Introduction}
Above the upper critical dimension $d=4$, for the Ising model with
nearest-neigbour interaction, the critical exponents assume their mean
field values \cite{aizenman:82, sokal:79}; $\alpha=0$ (finite specific
heat), $\beta=1/2$, $\gamma=1$, $\nu=1/2$, and the so called
hyperscaling law $d\nu=2-\alpha$ fails for $d>4$.  However, for
periodic boundary conditions most finite-size scaling properties near
the critical (inverse) temperature $K_c$ are well known.  For example,
the susceptibility behaves as $\chi \propto L^{5/2}$ for $d=5$, in
sharp contrast to $L^{\gamma/\nu}$ for $1<d<5$, with a logarithmic
correction for $d=4$. There is plenty of literature on $d>4$ for
periodic boundary conditions, to name but a few, see
e.g.~\cite{LBB:99, binder:08, jonesyoung:05, berche:08, brezin:85,
  pqpaper2, chendohm:00}.
  
Much less has been written on the subject of free boundary conditions
above the upper critical dimension, but see e.g.~\cite{boundarypaper,
  berche:12}. As can be seen from the references in those two papers
there has been some debate on whether the standard scaling picture,
saying that e.g. the susceptibility scales as $L^2$ for free boundary,
holds or whether an alternative theory proposing that it scales as
$L^{5/2}$ is correct. In \cite{boundarypaper} the current authors
simulated the 5-dimensional model on larger systems than previous
authors and found that the data supported the standard scaling
picture. In a reply \cite{berche:12} it was again suggested that the
alternative picture is correct and that the results of
\cite{boundarypaper} were due to too small systems, dominated by
finite size effects stemming from their large boundaries.

The purpose of this paper is two-fold. We have extended the
5-dimensional simulations with free boundary from \cite{boundarypaper}
to much larger systems, up to $L=160$, where the boundary vertices
make up less than $6.1\%$ of the system.  First, using the new data,
we give improved estimates of the critical energy, specific heat and
several other quantities at the critical point. Second, we compare how
well the standard scaling and the alternative theory fit our new large
system data, and discuss why, based on mathematical results on the
random cluster model, there are good reasons for expecting the
standard picture to be the correct one, as the data also suggests.

\section{Definitions and details}
For a given graph $G$ the Hamiltonian with interactions of unit
strength along the edges is $\mathcal{H}=-\sum_{ij} S_i S_j$ where the
sum is taken over the edges $ij$. As usual $K=1/k_BT$ is the
dimensionless inverse temperature and we denote the temperature
equilibrium mean by $\mean{\cdots}$. The susceptibility is defined as
$\chi = N \mean{m^2}$, where $m=(1/N)\sum_i S_i$ is the
magnetisation per spin, and the specific heat as $\heat=N
\left(\mean{U^2}-\mean{U}^2\right)$, where $U=(1/N)\sum_{ij}S_iS_j$ is
the energy per spin, and for short we write $\erg=\mean{U}$.

The underlying graph in question is an $L\times L\times L\times
L\times L$ grid graph with free boundary conditions, or equivalently,
the cartesian product of five paths on $L$ vertices. 

We have collected data for grid graphs of linear order $L=3$, $5$,
$7$, $11$, $15$, $19$, $23$, $31$, $39$, $47$, $55$, $63$, $72$, $96$,
$128$ and $160$. For $L=160$ we thus present data for systems on more
than 100 billion vertices.  States were generated with the Wolff
cluster method~\cite{wolff:89}. Between measurements, clusters were
updated until an expected $L^5$ spins were flipped.

For $3\le L\le 63$ we kept $64$ separate systems at nine couplings
$K=0.1139130$, $0.1139135$, $\ldots 0.1139170$.  For $L=72, 96$ we
used 48 separate systems, 116 for $L=128$ and 108 for $L=160$.  For
these larger systems the number of measurements were just short of
$30000$ for $L=72$ down to about 4000 for $L=160$. Means and standard
errors were estimated by exploiting the separate systems. For the
larger systems ($L\ge 72$), due to the comparably few measurements, we
also used bootstrapping on the entire data set for estimating standard
errors.  To double-check for equilibration problems we compared with
subsets of the data after rejecting early measurements.

The different couplings for $3\le L\le 63$ showed no discernible
difference in their scaling behaviour for $L\le 63$. Based on the
behaviour for $L\leq 63$ we have designated $K_c=0.1139150$, which is
a little higher than what we used in Ref.~\cite{boundarypaper} and
marginally lower than that used in e.g. Ref.~\cite{berche:12}. The
lion's share of sampling were then made at $0.1139150$ and for $L\ge
72$ we have measured only at $K_c$.

For all sizes we measured magnetisation and energy, storing their
moment sums. For $3\le L\le 63$ we also measured many properties
regarding the clusters that were generated and used for consistency
checks, and one of them will be shown in the Discussion section.

\subsection{Geometry and boundary effects}
A potentially important issue for systems with free boundary condition
is the size of the boundary, and in particular the fraction of
vertices on the boundary.  Of the $N=L^5$ vertices in the graph
$(L-2)^5$ are inner vertices and thus $L^5-(L-2)^5$ vertices sit on
the boundary. The fraction of boundary vertices is then
$1-(1-2/L)^5$. For $L=8$ this means that the boundary constitutes no
less than $76\%$ of all vertices. To continue, for $L=16$ the
boundary's share is $49\%$, for $L=32$ it is $28\%$, for $L=64$ it is
$15\%$, for $L=128$ it is $7.6\%$ and for $L=160$, our largest system
studied here, it makes up $6.1\%$.  So our largest system have a clear
minority of their vertices on the boundary.

Another important measure is the number of vertices with a given
minimum distance to the boundary.  If we consider the cube of side
$cL$ around the centre vertex of the cube, i.e. the set of vertices
with distance at least $\frac{cL}{2}$ to the boundary we find that it
contains a $c^5$ fraction of the $N$ vertices.  That means that at
least 50\% of the vertices are at a distance of at most $0.065L$ from
the boundary, for \emph{every} $L$.  Similarly, the central cube with
side $L/2$ contains just $3.1\%$ of the vertices of the cube.

This means that even in the limit the effect of vertices close to the
boundary will always be large, and that the vertices close to the
central vertex will also remain atypical for any property which depend
both on the distance to the boundary and the majority of the vertices
in the cube.  In particular we should expect such properties to be
bounded from above the corresponding values in the infinite system
thermodynamic limit, if they tend to decrease with the distance to the
boundary.

\section{Energy and specific heat}

\begin{figure}
  \begin{center}
    \includegraphics[width=0.483\textwidth]{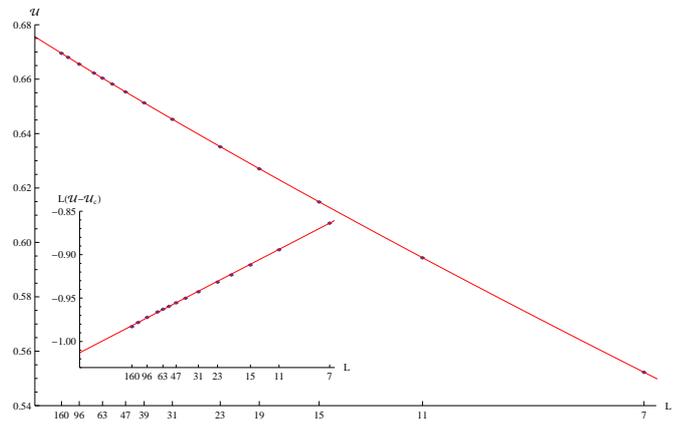}
  \end{center}
  \caption{(Colour on-line) Mean energy $\erg$ at $K_c$ versus $1/L$
    for $L=7$, $11$, $15$, $19$, $23$, $31$, $39$, $47$, $55$, $63$,
    $72$, $96$, $128$ and $160$. The red curve is $0.675647 - 1.013\,x
    + 0.395 \,x^{3/2}$ where $x=1/L$. Error bars are too small to be
    seen. See text for details. The inset shows the scaled energy
    $L(\erg-\erg_c)$ at $K_c$ versus $1/L^{1/2}$ for the same range of
    $L$.  Error bars are too small to be seen. The red curve $- 1.013
    + 0.395\,x$.  }\label{fig:u}
\end{figure}

It is known \cite{sokal:79} that in the limit ,for $d\geq 5$, the
specific heat, i.e. the energy variance, is bounded for all
temperatures, but the value is not known, and likewise for the
critical energy.  In Fig.~\ref{fig:u} the mean energy $\erg$ is shown
versus $1/L$.  The leading scaling term is here set to the order $1/L$
and the correction term to order $1/L^{3/2}$. This gave by far the
most stable coefficients of the fitted curve among the simple
exponents. We find that the best fitted curve is $0.675647(3) -
1.013(1)\,x + 0.395(1)\,x^{3/2}$, where $x=1/L$. The fit is excellent
down to $L=3$. The coefficients and their error estimates are here
based on the median and interquartile range of the coefficients when
deleting one of the data points from the fitting process.  In the
inset picture in Fig.~\ref{fig:u} we zoom into the plot by showing $L
(\erg-\erg_c)$ versus $1/L^{1/2}$ together with the line
$-1.013+0.395x$. The fit is vey good and hence we conclude that the
correction term is of the order $1/L^{1/2}$. Note that the error bars
in both plots are included but they are far too small to be seen at
this scale. The limit energy $\erg_c = \lim_{L\to\infty} \erg(K_c,L)
=0.675647(3)$ is only marginally larger than the value we gave
in~\cite{boundarypaper} which may be explained by the slightly smaller
$K_c$. Note that for free boundary conditions the limit is reached
from below whereas for periodic boundary conditions $\erg$ approaches
its limit from above, roughly as $\erg(K_c,L)-\erg_c \sim 5.5/L^{5/2}$
\cite{boundarypaper}.

\begin{figure}
  \begin{center}
    \includegraphics[width=0.483\textwidth]{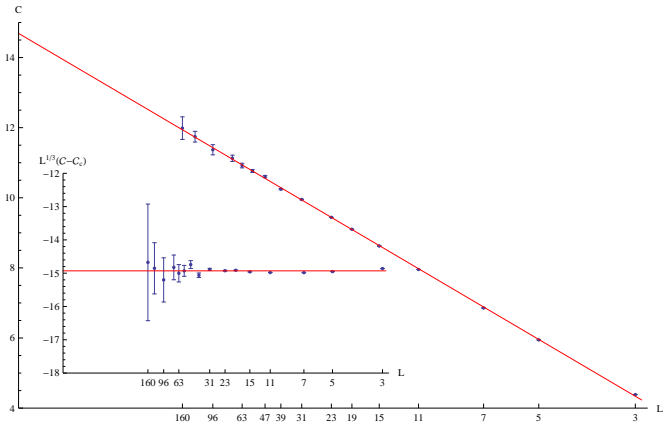}
  \end{center}
  \caption{(Colour on-line) Specific heat $\heat$ at $K_c$ versus
    $1/L^{1/3}$ for $L=3$, $5$, $7$, $11$, $15$, $19$, $23$, $31$,
    $39$, $47$, $55$, $63$, $72$, $96$, $128$ and $160$. The red line
    is $14.69 - 14.93\,x$ where $x=1/L^{1/3}$. The inset shows the
    scaled specific heat $L^{1/3}(\heat-\heat_c)$ at $K_c$ versus
    $1/L^{1/3}$ for the same range of $L$. The red line is the
    constant $-14.93$.  }\label{fig:c}
\end{figure}

For the specific heat we note that the error bars are noticeable, but
this is to be expected. It is not known at which rate $\heat(K_c,L)$
approaches its asymptotic value $\heat_c$ but judging from the
excellent line-up of the points in Fig.~\ref{fig:c} it appears
$\heat(K_c,L)-\heat_c \propto 1/L^{1/3}$. This is the only time we see
a $1/3$ in a scaling exponent for free boundary conditions and we have
no theoretical basis for it. In Fig.~\ref{fig:c} we show
$\heat(K_c,L)$ versus $1/L^{1/3}$ and the line $14.69(1) - 14.93(2)x$,
where $x=1/L^{1/3}$. As before the error estimates of the coefficients
are based on the variability of a fitted curve after deleting one of
the points. We estimate thus that $\heat_c =
\lim_{L\to\infty}\heat(K_c,L) = 14.69(1)$. The inset picture
of Fig.~\ref{fig:c} zoom into the correction term by plotting
$L^{1/3}(\heat-\heat_c)$ versus $1/L^{1/3}$ together with the constant
line $-14.93$. The error bars now become quite noticeable, especially
for the larger $L$. There is no clear trend upwards or downwards in
the data points which suggests that any further corrections to scaling
must be truly negligible.

\section{Magnetisation and susceptibility}
\begin{figure}
  \begin{center}
    \includegraphics[width=0.483\textwidth]{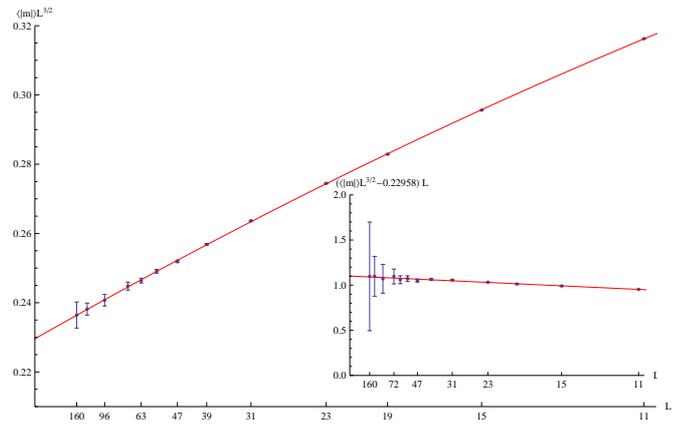}
  \end{center}
  \caption{(Colour on-line) Modulus magnetisation $\langle |m| \rangle
    L^{3/2}$ at $K_c$ versus $1/L$ for $L=11$, $15$, $19$, $23$, $31$,
    $39$, $47$, $55$, $63$, $72$, $96$, $128$ and $160$. The red curve
    is $0.22958 + 1.101\,x - 1.63 \,x^2$, where $x=1/L$.  The inset
    shows the scaled magnetisation $(\langle |m| \rangle
    L^{3/2}-0.22958) L$ at $K_c$ versus $1/L$ for the same range of
    $L$. The red line is $1.101 - 1.63\,x$.
  }\label{fig:m}
\end{figure}

The scaling of the modulus of the magnetisation $\langle |m|\rangle$
for free boundary conditions is very different from that of periodic
boundary conditions. In the first case we find $\langle |m| \rangle
\propto L^{-3/2}$ whereas in the second it is well-known that $\langle
|m| \rangle \propto L^{-5/4}$. In the free boundary case we note the
need for correction to scaling. Indeed, if we want perfectly fitted
curves down to $L=3$ we need two correction terms. We have instead
chosen to ignore $L\le 7$ and stay with just one correction term for
the remaining $13$ points. In Fig.~\ref{fig:m} we show $\langle |m|
\rangle L^{3/2}$ versus $1/L$ for $11\le L\le 160$ together with the
curve $0.22958(6) + 1.101(3)\,x - 1.63(3) \,x^2$.  We test the fit of
this curve by zooming into the picture and instead show $(\langle |m|
\rangle L^{3/2} - 0.22958) L$ which then should be well fitted by the line
$1.101 - 1.63\,x$. As the inset of Fig~\ref{fig:m} shows, it is
and we conclude that to leading order $\langle |m| \rangle \sim
0.22958(6) L^{-3/2}$. However, the error bars for the largest systems
are now quite pronounced.

\begin{figure}
  \begin{center}
    \includegraphics[width=0.483\textwidth]{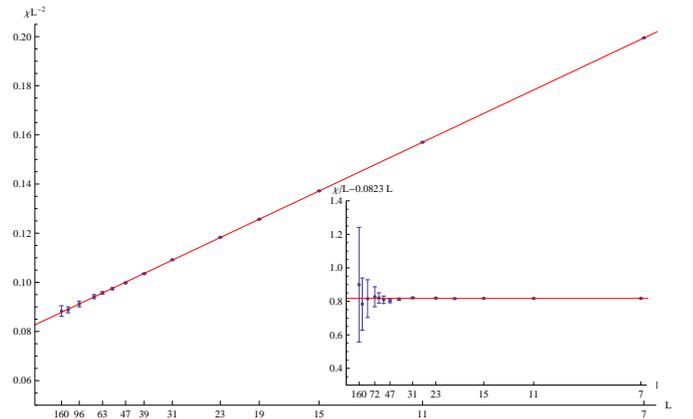}
  \end{center}
  \caption{(Colour on-line) Normalised susceptibility $\chi/L^2$ at
    $K_c$ versus $1/L$ for $L=7$, $11$, $15$, $19$, $23$, $31$, $39$,
    $47$, $55$, $63$, $72$, $96$, $128$ and $160$. The red line is
    $0.08269 + 0.8174\,x$, where $x=1/L$. The inset shows the scaled
    and normalised susceptibility $(\chi/L^2-0.08269) L$ at $K_c$
    versus $1/L$ for the same range of $L$. The red line is the
    constant $0.8174$. }\label{fig:x}
\end{figure}

As we mentioned above the susceptibility $\chi=N \langle m^2\rangle$
scales to leading order as $\chi\propto L^{5/2}$ for periodic boundary
conditions but it is not known what the corresponding order is for
free boundary conditions. We find here, as in
Ref.~\cite{boundarypaper}, that $\chi\propto L^2$ is by far the best
scaling rule. In Fig.~\ref{fig:x} we show $\chi/L^2$ versus $1/L$ for
$7\le L\le 160$ together with the line $0.08269(2) +
0.8174(3)\,x$. The coefficients were determined after excluding $L\le
5$ from the fitting process. Had we included the two smallest systems
an extra correction to scaling term would have been required. Again we
zoom in and the inset of Fig.~\ref{fig:x} shows $(\chi/L^2-0.08269) L$
versus $1/L$ together with the constant line $0.8174$. Though the
error bars are quite big for the largest systems the fit is quite
acceptable.  In short we find $\chi(K_c) \sim 0.08269(2) L^{2}$. We
might add that the corresponding expression for periodic boundary is
not known exactly but we suggested recently \cite{pqpaper2} that $\chi
\sim 1.742 L^{5/2}$.

\section{Susceptibility compared to $L^{5/2}$}
It has been suggested \cite{Berche2012115} that $L^{2}$ is in fact not
the correct scaling for the susceptibility. The authors of
\cite{Berche2012115} claim that the correct scaling should be
$L^{5/2}$, as for the case with cyclic boundary conditions, and that
the exponent previously found by us, and other authors, are based on
either finite size effects due to too many boundary vertices in small
systems, for simulation studies, or incomplete theory.  In our
previous work the boundary did indeed contain a large fraction of the
system's vertices but in our current study this fraction has been
reduced to a lower value than in any previous study, including the
truncated systems used in \cite{Berche2012115}.

\begin{figure}
  \begin{center}
    \includegraphics[width=0.483\textwidth]{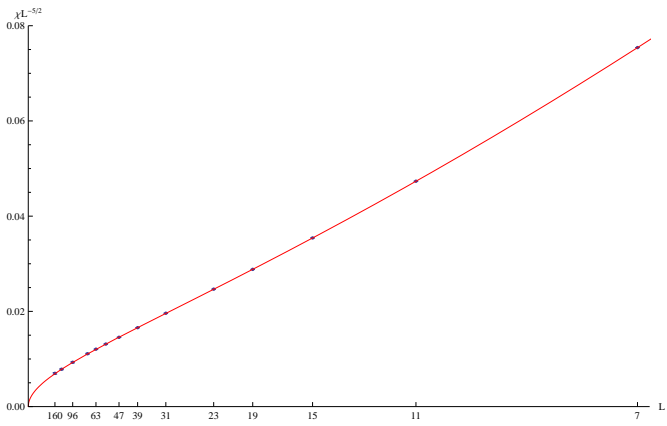}
  \end{center}
  \caption{(Colour on-line) Normalised susceptibility $\chi/L^{5/2}$
    at $K_c$ versus $1/L$ for $L=7$, $11$, $15$, $19$, $23$, $31$,
    $39$, $47$, $55$, $63$, $72$, $96$, $128$ and $160$. The red curve
    is $0.085\, x^{0.51} + 0.82\, x^{1.51}$, where $x=1/L$.
  }\label{fig:x52}
\end{figure}

To avoid implicit bias in our scaling of the susceptibility to we can
also test the ratio $\chi/L^{5/2}$.  If the claims of
\cite{Berche2012115} are correct this quantity should converge to a
finite non-zero limit, at least for large enough systems, and if the
standard scaling is correct it should to leading order converge to 0
as $L^{-0.5}$.  We make a scaling ansatz $c_0+ c_1 \,x^{\lambda_1} +
c_2 \,x^{\lambda_2}$, where $x=1/L$, and let Mathematica find the five
free parameters using a least squares fit, after excluding $L=3,5$.
As usual, we let each remaining point be deleted in turn from the
fitting data to obtain error bars of the parameters. We find on
average the curve $0.0000(4) + 0.085(9)\, x^{0.51(3)} + 0.820(7)\,
x^{1.51(3)}$. Clearly the parameters we estimated above for $\chi/L^2$
falls inside these estimates, though the error bars are a magnitude
larger here. Using the middle point values we plot the curve together
with the data points in Fig.~\ref{fig:x52}.  The data for large
systems is clearly consistent with the standard scaling, even for an
unrestricted data fitting like this.

\section{Fourth moment and kurtosis}
\begin{figure}
  \begin{center}
    \includegraphics[width=0.483\textwidth]{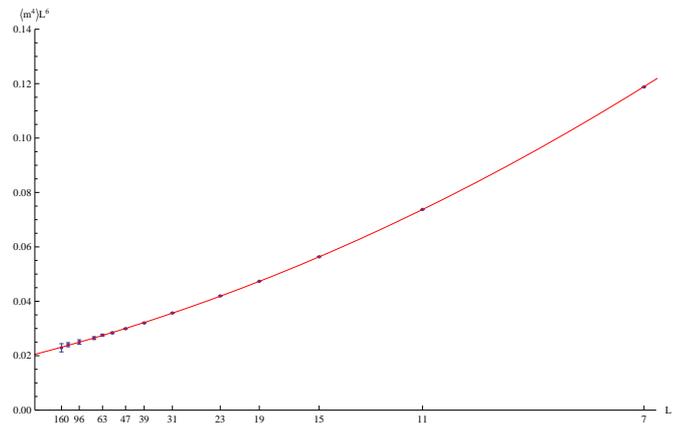}
  \end{center}
  \caption{(Colour on-line) Normalised fourth moment $\langle m^4
    \rangle L^6$ at $K_c$ versus $1/L$ for $L=7$, $11$, $15$, $19$,
    $23$, $31$, $39$, $47$, $55$, $63$, $72$, $96$, $128$ and
    $160$. The red curve is $0.02051 + 0.4045\,x + 1.989 \,x^2$, where
    $x=1/L$.  }\label{fig:m4}
\end{figure}

Our final property of interest is the fourth moment of the
magnetisation at $K_c$.  We find that the fourth moment of the
magnetisation scales as $\langle m^4\rangle\propto L^{-6}$. The
general rule would then be $\langle |m^k|\rangle \propto L^{-3k/2}$
whereas the corresponding rule for periodic boundary conditions is
$\langle |m^k|\rangle \propto L^{-5k/4}$. Proceeding in the same
manner as before, we plot the normalised fourth moment's behaviour as
$\langle m^4\rangle L^6$ versus $1/L$ in Fig.~\ref{fig:m4} together
with the estimated polynomial $0.02051(3) + 0.4045(8)\,x + 1.989(4)
\,x^2$. We excluded $L=3$ from the coefficient estimates. To leading
order we thus find $\langle m^4\rangle \sim 0.02051(3) L^{-6}$.

\begin{figure}
  \begin{center}
    \includegraphics[width=0.483\textwidth]{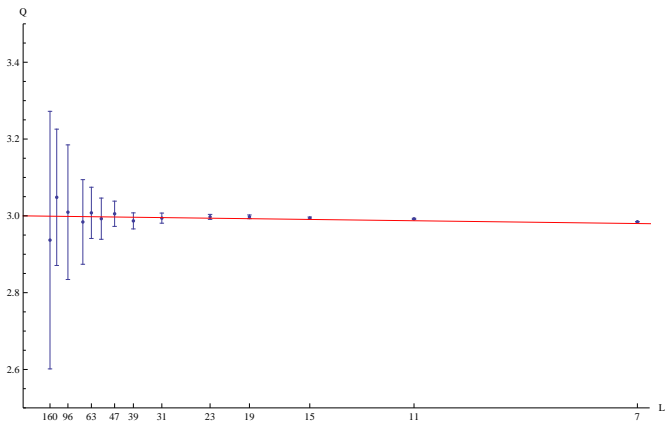}
  \end{center}
  \caption{(Colour on-line) Kurtosis $\langle m^4\rangle / \langle
    m^2\rangle^2$ at $K_c$ versus $1/L$ for the same range of $L$ as
    in Fig.~\ref{fig:m4}. The red line is $3 - 0.14\,x$, where $x=1/L$.
  }\label{fig:Q}
\end{figure}

The moment ratio $Q=\langle m^4\rangle / \langle m^2\rangle^2$, or
kurtosis, indicates the shape of the underlying magnetisation
distribution. In Fig.~\ref{fig:Q} we show the kurtosis versus $1/L$
and the line $3 - 0.14\,x$. The error bars are based on the formula for
the error of a quotient, $\dd (x/y^2)$.
The line is based on the coefficient estimates for $m^2$ and $m^4$
above by taking the quotient of their respective series expansions in
the standard fashion.
Inserting the coefficients and their error estimates gives the limit
$\langle m^4\rangle / \langle m^2\rangle^2 \to 3.000(6)$ which is the
characteristic value of a gaussian distribution. Recall that for
periodic boundary conditions the kurtosis at $K_c$ takes the
asymptotic value $\Gamma(1/4)^4/2\pi^2 = 2.1884\ldots$, see
Refs.~\cite{brezin:85, pqpaper2}

\section{Kurtosis at an effective critical point}
\begin{figure}
  \begin{center}
    \includegraphics[width=0.483\textwidth]{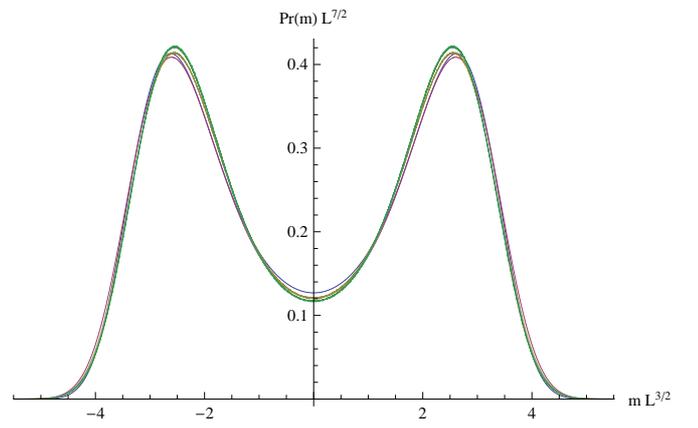}
  \end{center}
  \caption{(Colour on-line) Scaled magnetisation distributions $\Pr(m)
    L^{7/2}$ versus $m L^{3/2}$ at $K_c(L)$ (see text) for $L=4$, $6$,
    $8$ and $10$.  }\label{fig:mdist2}
\end{figure}

\begin{figure}
  \begin{center}
    \includegraphics[width=0.483\textwidth]{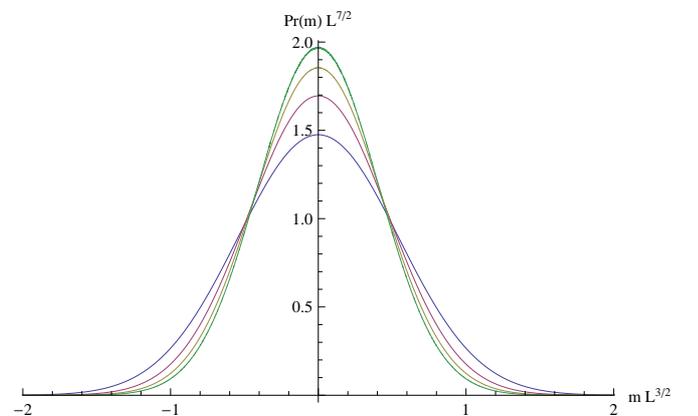}
  \end{center}
  \caption{(Colour on-line) Scaled magnetisation distributions $\Pr(m)
    L^{7/2}$ versus $m L^{3/2}$ at $K_c$ for $L=4$, $6$, $8$ and $10$
    (increasing at $y$-axis). Data from Ref.~\cite{boundarypaper}.
  }\label{fig:mdist}
\end{figure}

\begin{figure}
  \begin{center}
    \includegraphics[width=0.483\textwidth]{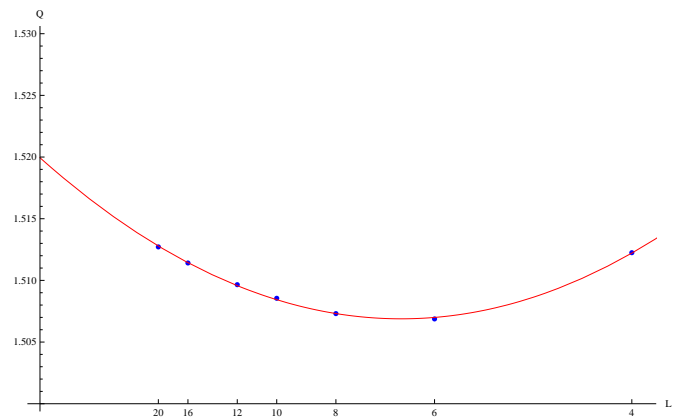}
  \end{center}
  \caption{(Colour on-line) Kurtosis $Q=\langle m^4\rangle / \langle
    m^2\rangle^2$ at $K_c(L)$ (see text) versus $1/L$ for $L=4$, $6$,
    $8$, $10$, $12$, $16$, $20$. The red curve $1.52 - 0.17\,x +
    0.56 \,x^2$, where $x=1/L$.  }\label{fig:Q2}
\end{figure}

That the kurtosis takes the asymptotical value $3$ at $K_c$ does of
course not mean that the kurtosis converges to $3$ for every sequence of
temperatures $K_c(L)$ that has $K_c$ as limit. We exemplify this by
reexamining some of our data used in Ref.~\cite{boundarypaper} where
we relied on extremely detailed data on a wide temperature range for
$4\le L \le 20$. For $L=4,6,8,10$ we also have magnetisation
distributions. Let us say that $K_c(L)$ is the point where the
variance of the modulus magnetisation, i.e. $\bar\chi =
N\left(\mean{m^2} - \mean{|m|}^2\right)$, takes its maximum value. The
distribution is here at its widest and on the threshold of breaking up
into two parts, see Fig.~\ref{fig:mdist2} where we show a scaled
distribution at $K_c(L)$ for $L=4,6,8,10$, in stark contrast to the
distribution at $K_c$ of Fig.~\ref{fig:mdist}. Measuring the kurtosis
at this point produces Fig.~\ref{fig:Q2} which shows $Q(K_c(L), L)$
versus $1/L$ and a fitted 2nd degree polynomial which suggests the
limit $1.520$. The absence of error bars is due to the method by which
the original data were produced. However, we expect the error to be
smaller than the plotted points. In fact, repeating this exercise for
periodic boundary conditions suggests the limit $1.517$. There is a
distinct possibility that these two limits are in fact the same, but
that would require high-resolution data for larger systems to resolve
than is at our disposal.  In any case this subject falls outside the
scope of this paper.

\section{Discussion and Conclusions}
As we have seen the sampled data for cubes up to side $L=160$ agree
well with the standard scaling picture for free boundary conditions,
and e.g. recent long series expansions \cite{PhysRevE.85.021105} also
appears to favour this version, nontheless without a rigorous bound
for the rate of convergence a simulation study is always open to the
claim that it is dominated by finite size effects.

However, the last decade has seen a number of rigorous mathematical
results on the behaviour of the random-cluster model which leads us to
believe that the standard scaling is indeed the right one. The
Fortuin-Kasteleyn random-cluster model has a parameter $q$ which
governs the properties of the model. We'll refer the reader to
\cite{grimmett2004random} for more details and history.  We recall
that for $q=1$ the model is the standard bond percolation model, for
$q=2$ it is equivalent to the Ising model, and in the limit
$q\rightarrow 0$ we get the uniform random spanning tree, or the
uniform spanning forest, depending on the parameter $p$.

For the random spanning tree on the $d$-dimensional lattice with
different boundary conditions Pemantle \cite{MR1127715} begun a study
which related it to the loop erased random walk and demonstrated a
strong dependency on both the dimension and boundary condition. In
later papers \cite{MR2172682, MR2391250,MR2496437} these results were
refined to show, among other things, that for large enough $d$ that
for two points in a grid with side $L$ and free boundary the distance
between the two points within the tree scale as $L^2$, but that for
the torus of side $L$ the distance scales as $L^{d/2}$.

Coming to the case $q=1$, Aizenman \cite{aizenman:97} studied the
behaviour of the largest crossing clusters, i.e clusters which contain
vertices on opposite sides of the box, in percolation on grids in
different dimensions. Among other things he conjectured that at the
critical point $p_c$, for all large enough dimensions $d$ the largest
cluster in the case with free boundary should scale as $L^4$, and for
the torus, or cyclic boundary, it should scale as $L^{2d/3}$.  This
conjecture was proven in \cite{HH:07,HH:11}, and so we know that for
percolation the boundary condition have a large and non-vanishing
effect on the size of the largest clusters at the critical point. It
is also important to point out that the results from
\cite{HH:07,HH:11} are for the scaling \emph{exactly} at the critical
point $p_c$. For other sequences of points converging to $p_c$
different scaling behaviours can appear.

The reason for the drastic difference between the torus and free
boundary case here is that for large $d$ the clusters in the model
become much more spread out than in low dimensions. For $d=2$ clusters
at $p_c$ are always finite and have a boundary which in the scaling
limit is very close to a brownian motion \cite{MR1879816}.  For a
finite box of side $L$, with free boundary the number of crossing
clusters, has a finite mean, bounded as $L$ grows, and the probability
that there are more than $k$ such clusters is less than $\exp(-a_1
k^2)$, for some positive constant $a_1$ \cite{aizenman:97}. For a box
with free boundary in $d>6$, the critical dimension for $q=1$, the
number of crossing clusters is at least $a_2 L^{d-6}$, for some
constant $a_2$ \cite{aizenman:97}. Further, the largest cluster and a
positive proportion of the crossing clusters have size proportional to
$L^4$, independently of $d$. The clusters are here of much lower
dimension relative to that of the lattice and much more tree like in
their structure than for $d=2$. If we instead consider a torus with
$d>6$ the large number of these more expanding clusters lead them to
connect up with other clusters, which would have been separate in the
free boundary case, and this merging leads to maximum clusters that
are vastly larger than in the free boundary case.

The current authors expect a picture similar to that for $q=1$ to hold
for the Ising case $q=2$ as well. Since the susceptibility in the
Ising model is proportional to the average cluster size in the
random-cluster model this leads to a prediction of $L^2$ as the
correct scaling for the free boundary case and $L^{5/2}$ for cyclic
boundaries, for $d=5$.  This would also lead us to expect the largest
clusters in the free boundary case to scale as $L^2$. For our small to
medium sized systems we collected the size of the cluster containing
the central vertex of our cubes, a property which was also considered
in \cite{Berche2012115} for truncated systems.

\begin{figure}
  \begin{center}
    \includegraphics[width=0.483\textwidth]{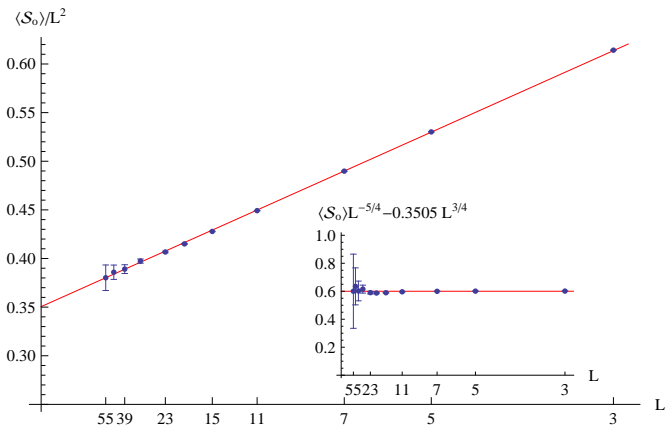}
  \end{center}
\caption{(Colour on-line) The normalised average size of the cluster
  containing the centre vertex, $\langle S_0\rangle/L^2$, plotted
  versus $1/L^{3/4}$ for $L=3$, $5$, $7$, $11$, $15$, $19$, $23$,
  $31$, $39$, $47$, $55$. The red line is $0.3505+0.600x$ where
  $x=1/L^{3/4}$.  The inset shows the scaled and normalised cluster
  size $\langle S_0\rangle L^{-5/4}-0.3505 L^{3/4}$ versus $1/L$. The
  red line is the constant $0.600$.}\label{fig:s0}
\end{figure}

In Fig.~\ref{fig:s0} we show the normalised mean cluster size
$\langle S_0\rangle/L^2$ for $3\le L\le 55$ as used in
\cite{Berche2012115}, together with a linear function estimated to be
$0.3505(2)+0.600(1) x$.  The inset shows the zoomed-in version
$\langle S_0\rangle L^{-5/4}-0.3505 L^{3/4}$ versus $1/L$ which then
should essentially take the constant value $0.600$, the red line. As
we can see we have an excellent fit to the prediction that this
cluster size should scale as $L^2$.

Aizenman's prediction \cite{aizenman:97} of $L^{2d/3}$ as the correct
scaling for percolation on the torus, for high $d$, came from a
comparison with the Erd\"os-Renyi random graph on $N$ vertices, for
which the largest connected component at the critical probability,
scales as $N^{2/3}$.  In \cite{5dart} we follow this analogy further
by comparing the largest cluster for the random-cluster model on
5-dimensional tori with the detailed rigorous results on the
random-cluster model for complete graphs from \cite{luczak:06}, and a
good agreement is found.

To conclude, we find that both the data from our simulations and the
current mathematical result for the random-cluster model gives good
support for the standard scaling picture for the Ising model with free
boundary conditions, as well as a framework predicting further
properties for the case with cyclic boundary.

\section{Acknowledgements}
The simulations were performed on resources provided by the Swedish
National Infrastructure for Computing (SNIC) at High Performance
Computing Center North (HPC2N) and at Chalmers Centre for
Computational Science and Engineering (C3SE).


\begin{thebibliography}{25}
\expandafter\ifx\csname natexlab\endcsname\relax\def\natexlab#1{#1}\fi
\expandafter\ifx\csname bibnamefont\endcsname\relax
  \def\bibnamefont#1{#1}\fi
\expandafter\ifx\csname bibfnamefont\endcsname\relax
  \def\bibfnamefont#1{#1}\fi
\expandafter\ifx\csname citenamefont\endcsname\relax
  \def\citenamefont#1{#1}\fi
\expandafter\ifx\csname url\endcsname\relax
  \def\url#1{\texttt{#1}}\fi
\expandafter\ifx\csname urlprefix\endcsname\relax\def\urlprefix{URL }\fi
\providecommand{\bibinfo}[2]{#2}
\providecommand{\eprint}[2][]{\url{#2}}

\bibitem[{\citenamefont{Aizenman}(1982)}]{aizenman:82}
\bibinfo{author}{\bibfnamefont{M.}~\bibnamefont{Aizenman}},
  \bibinfo{journal}{Comm. Math. Phys.} \textbf{\bibinfo{volume}{86}},
  \bibinfo{pages}{1} (\bibinfo{year}{1982}), ISSN \bibinfo{issn}{0010-3616}.

\bibitem[{\citenamefont{Sokal}(1979)}]{sokal:79}
\bibinfo{author}{\bibfnamefont{A.~D.} \bibnamefont{Sokal}},
  \bibinfo{journal}{Phys. Lett. A} \textbf{\bibinfo{volume}{71}},
  \bibinfo{pages}{451} (\bibinfo{year}{1979}).

\bibitem[{\citenamefont{Luijten et~al.}(1999)\citenamefont{Luijten, Binder, and
  Bl\"ote}}]{LBB:99}
\bibinfo{author}{\bibfnamefont{E.}~\bibnamefont{Luijten}},
  \bibinfo{author}{\bibfnamefont{K.}~\bibnamefont{Binder}}, \bibnamefont{and}
  \bibinfo{author}{\bibfnamefont{H.}~\bibnamefont{Bl\"ote}},
  \bibinfo{journal}{Eur. Phys. J. B} \textbf{\bibinfo{volume}{9}},
  \bibinfo{pages}{289} (\bibinfo{year}{1999}).

\bibitem[{\citenamefont{Binder}(2008)}]{binder:08}
\bibinfo{author}{\bibfnamefont{K.}~\bibnamefont{Binder}},
  \bibinfo{journal}{Eur. Phys. J. B} \textbf{\bibinfo{volume}{64}},
  \bibinfo{pages}{307} (\bibinfo{year}{2008}), ISSN \bibinfo{issn}{1434-6028}.

\bibitem[{\citenamefont{Jones and Young}(2005)}]{jonesyoung:05}
\bibinfo{author}{\bibfnamefont{J.~L.} \bibnamefont{Jones}} \bibnamefont{and}
  \bibinfo{author}{\bibfnamefont{A.~P.} \bibnamefont{Young}},
  \bibinfo{journal}{Phys. Rev. B} \textbf{\bibinfo{volume}{71}},
  \bibinfo{pages}{174438} (\bibinfo{year}{2005}).

\bibitem[{\citenamefont{Berche et~al.}(2008)\citenamefont{Berche, Chatelain,
  Dhall, Kenna, Low, and Walter}}]{berche:08}
\bibinfo{author}{\bibfnamefont{B.}~\bibnamefont{Berche}},
  \bibinfo{author}{\bibfnamefont{C.}~\bibnamefont{Chatelain}},
  \bibinfo{author}{\bibfnamefont{C.}~\bibnamefont{Dhall}},
  \bibinfo{author}{\bibfnamefont{R.}~\bibnamefont{Kenna}},
  \bibinfo{author}{\bibfnamefont{R.}~\bibnamefont{Low}}, \bibnamefont{and}
  \bibinfo{author}{\bibfnamefont{J.-C.} \bibnamefont{Walter}},
  \bibinfo{journal}{J. Stat. Mech.} \textbf{\bibinfo{volume}{2008}},
  \bibinfo{pages}{P11010} (\bibinfo{year}{2008}).

\bibitem[{\citenamefont{Brezin and Zinn-Justin}(1985)}]{brezin:85}
\bibinfo{author}{\bibfnamefont{E.}~\bibnamefont{Brezin}} \bibnamefont{and}
  \bibinfo{author}{\bibfnamefont{J.}~\bibnamefont{Zinn-Justin}},
  \bibinfo{journal}{Nucl. Phys. B} \textbf{\bibinfo{volume}{257}},
  \bibinfo{pages}{867} (\bibinfo{year}{1985}).

\bibitem[{\citenamefont{Lundow and Rosengren}(2013)}]{pqpaper2}
\bibinfo{author}{\bibfnamefont{P.~H.} \bibnamefont{Lundow}} \bibnamefont{and}
  \bibinfo{author}{\bibfnamefont{A.}~\bibnamefont{Rosengren}},
  \bibinfo{journal}{Phil. Mag.} \textbf{\bibinfo{volume}{93}},
  \bibinfo{pages}{1755} (\bibinfo{year}{2013}).

\bibitem[{\citenamefont{Chen and Dohm}(2000)}]{chendohm:00}
\bibinfo{author}{\bibfnamefont{X.~S.} \bibnamefont{Chen}} \bibnamefont{and}
  \bibinfo{author}{\bibfnamefont{V.}~\bibnamefont{Dohm}},
  \bibinfo{journal}{Phys. Rev. E} \textbf{\bibinfo{volume}{63}},
  \bibinfo{pages}{016113} (\bibinfo{year}{2000}).

\bibitem[{\citenamefont{Lundow and Markstr\"om}(2011)}]{boundarypaper}
\bibinfo{author}{\bibfnamefont{P.~H.} \bibnamefont{Lundow}} \bibnamefont{and}
  \bibinfo{author}{\bibfnamefont{K.}~\bibnamefont{Markstr\"om}},
  \bibinfo{journal}{Nucl. Phys. B} \textbf{\bibinfo{volume}{845}},
  \bibinfo{pages}{120 } (\bibinfo{year}{2011}).

\bibitem[{\citenamefont{Berche et~al.}(2012{\natexlab{a}})\citenamefont{Berche,
  Kenna, and Walter}}]{berche:12}
\bibinfo{author}{\bibfnamefont{B.}~\bibnamefont{Berche}},
  \bibinfo{author}{\bibfnamefont{R.}~\bibnamefont{Kenna}}, \bibnamefont{and}
  \bibinfo{author}{\bibfnamefont{J.-C.} \bibnamefont{Walter}},
  \bibinfo{journal}{Nucl. Phys. B} \textbf{\bibinfo{volume}{865}},
  \bibinfo{pages}{115 } (\bibinfo{year}{2012}{\natexlab{a}}).

\bibitem[{\citenamefont{Wolff}(1989)}]{wolff:89}
\bibinfo{author}{\bibfnamefont{U.}~\bibnamefont{Wolff}},
  \bibinfo{journal}{Phys. Rev. Lett} \textbf{\bibinfo{volume}{62}},
  \bibinfo{pages}{361} (\bibinfo{year}{1989}).

\bibitem[{\citenamefont{Berche et~al.}(2012{\natexlab{b}})\citenamefont{Berche,
  Kenna, and Walter}}]{Berche2012115}
\bibinfo{author}{\bibfnamefont{B.}~\bibnamefont{Berche}},
  \bibinfo{author}{\bibfnamefont{R.}~\bibnamefont{Kenna}}, \bibnamefont{and}
  \bibinfo{author}{\bibfnamefont{J.-C.} \bibnamefont{Walter}},
  \bibinfo{journal}{Nucl. Phys. B} \textbf{\bibinfo{volume}{865}},
  \bibinfo{pages}{115 } (\bibinfo{year}{2012}{\natexlab{b}}).

\bibitem[{\citenamefont{Butera and Pernici}(2012)}]{PhysRevE.85.021105}
\bibinfo{author}{\bibfnamefont{P.}~\bibnamefont{Butera}} \bibnamefont{and}
  \bibinfo{author}{\bibfnamefont{M.}~\bibnamefont{Pernici}},
  \bibinfo{journal}{Phys. Rev. E} \textbf{\bibinfo{volume}{85}},
  \bibinfo{pages}{021105} (\bibinfo{year}{2012}).

\bibitem[{\citenamefont{Grimmett}(2004)}]{grimmett2004random}
\bibinfo{author}{\bibfnamefont{G.}~\bibnamefont{Grimmett}},
  \emph{\bibinfo{title}{The random-cluster model}}
  (\bibinfo{publisher}{Springer}, \bibinfo{year}{2004}).

\bibitem[{\citenamefont{Pemantle}(1991)}]{MR1127715}
\bibinfo{author}{\bibfnamefont{R.}~\bibnamefont{Pemantle}},
  \bibinfo{journal}{Ann. Probab.} \textbf{\bibinfo{volume}{19}},
  \bibinfo{pages}{1559} (\bibinfo{year}{1991}).

\bibitem[{\citenamefont{Benjamini and Kozma}(2005)}]{MR2172682}
\bibinfo{author}{\bibfnamefont{I.}~\bibnamefont{Benjamini}} \bibnamefont{and}
  \bibinfo{author}{\bibfnamefont{G.}~\bibnamefont{Kozma}},
  \bibinfo{journal}{Comm. Math. Phys.} \textbf{\bibinfo{volume}{259}},
  \bibinfo{pages}{257} (\bibinfo{year}{2005}).

\bibitem[{\citenamefont{Schweinsberg}(2008)}]{MR2391250}
\bibinfo{author}{\bibfnamefont{J.}~\bibnamefont{Schweinsberg}},
  \bibinfo{journal}{J. Theoret. Probab.} \textbf{\bibinfo{volume}{21}},
  \bibinfo{pages}{378} (\bibinfo{year}{2008}).

\bibitem[{\citenamefont{Schweinsberg}(2009)}]{MR2496437}
\bibinfo{author}{\bibfnamefont{J.}~\bibnamefont{Schweinsberg}},
  \bibinfo{journal}{Probab. Theory Related Fields}
  \textbf{\bibinfo{volume}{144}}, \bibinfo{pages}{319} (\bibinfo{year}{2009}).

\bibitem[{\citenamefont{Aizenman}(1997)}]{aizenman:97}
\bibinfo{author}{\bibfnamefont{M.}~\bibnamefont{Aizenman}},
  \bibinfo{journal}{Nucl. Phys. B} \textbf{\bibinfo{volume}{485}},
  \bibinfo{pages}{551} (\bibinfo{year}{1997}).

\bibitem[{\citenamefont{Heydenreich and van~der Hofstad}(2007)}]{HH:07}
\bibinfo{author}{\bibfnamefont{M.}~\bibnamefont{Heydenreich}} \bibnamefont{and}
  \bibinfo{author}{\bibfnamefont{R.}~\bibnamefont{van~der Hofstad}},
  \bibinfo{journal}{Comm. Math. Phys.} \textbf{\bibinfo{volume}{270}},
  \bibinfo{pages}{335} (\bibinfo{year}{2007}).

\bibitem[{\citenamefont{Heydenreich and van~der Hofstad}(2011)}]{HH:11}
\bibinfo{author}{\bibfnamefont{M.}~\bibnamefont{Heydenreich}} \bibnamefont{and}
  \bibinfo{author}{\bibfnamefont{R.}~\bibnamefont{van~der Hofstad}},
  \bibinfo{journal}{Probab. Theory Related Fields}
  \textbf{\bibinfo{volume}{149}}, \bibinfo{pages}{397} (\bibinfo{year}{2011}).

\bibitem[{\citenamefont{Smirnov and Werner}(2001)}]{MR1879816}
\bibinfo{author}{\bibfnamefont{S.}~\bibnamefont{Smirnov}} \bibnamefont{and}
  \bibinfo{author}{\bibfnamefont{W.}~\bibnamefont{Werner}},
  \bibinfo{journal}{Math. Res. Lett.} \textbf{\bibinfo{volume}{8}},
  \bibinfo{pages}{729} (\bibinfo{year}{2001}).

\bibitem[{\citenamefont{Lundow and Markstr\"om}()}]{5dart}
\bibinfo{author}{\bibfnamefont{P.~H.} \bibnamefont{Lundow}} \bibnamefont{and}
  \bibinfo{author}{\bibfnamefont{K.}~\bibnamefont{Markstr\"om}},
  \bibinfo{note}{arXiv:1408.2155}.

\bibitem[{\citenamefont{Luczak and Luczak}(2006)}]{luczak:06}
\bibinfo{author}{\bibfnamefont{M.~J.} \bibnamefont{Luczak}} \bibnamefont{and}
  \bibinfo{author}{\bibfnamefont{T.}~\bibnamefont{Luczak}},
  \bibinfo{journal}{Random Struct. Algorithms} \textbf{\bibinfo{volume}{28}},
  \bibinfo{pages}{215} (\bibinfo{year}{2006}).

\end{thebibliography}

\end{document}